# Self-supervised Fine-tuning for Correcting Super-Resolution Convolutional Neural Networks

Alice Lucas, Santiago Lopez-Tapia, Rafael Molina and Aggelos K. Katsaggelos.

*Abstract*—While Convolutional Neural Networks (CNNs) trained for image and video super-resolution (SR) regularly achieve new state-of-the-art performance, they also suffer from significant drawbacks. One of their limitations is their lack of robustness to unseen image formation models during training. Other limitations include the generation of artifacts and hallucinated content when training Generative Adversarial Networks (GANs) for SR. While the Deep Learning literature focuses on presenting new training schemes and settings to resolve these various issues, we show that one can avoid training and correct for SR results with a fully self-supervised fine-tuning approach. More specifically, at test time, given an image and its known image formation model, we fine-tune the parameters of the trained network and iteratively update them using a data fidelity loss. We apply our fine-tuning algorithm on multiple image and video SR CNNs and show that it can successfully correct for a sub-optimal SR solution by entirely relying on internal learning at test time. We apply our method on the problem of fine-tuning for unseen image formation models and on removal of artifacts introduced by GANs.

## I. Introduction

In the past decade, the application of Convolutional Neural Networks (CNNs) for solving inverse imaging problems has gained considerable popularity [2]. With the use of deep perceptual losses when training CNNs for super-resolution (SR), such as feature and adversarial losses, CNNs are capable of producing SR estimates of previously unseen perceptual quality ([3], [4], [5]). However, most current DL publications still suffer from major limitations. One such limitation is that a CNN trained for a specific image recovery problem will not generalize well to a test data point whose image formation model differs from the model established when synthesizing the training dataset. For example in the case of super-resolution, CNNs trained for a specific scale factor do not generalize well to new scale factors at test time. Inputting a low-resolution (LR) image to an SR CNN that has been downsampled by a different scale factor may result in unnatural-looking images with unpleasant artifacts at the output layer of the CNN. Consider also the case in which blur is applied to a high-resolution (HR) image prior to being downsampled and provided as input to a trained CNN. If the CNN was not trained to expect such blur at the LR input then the super-resolution will fail [6].

In the Deep Learning literature, the standard procedure to improving generalization capabilities of SR CNNs consists of modifying the training scheme and designing network architectures that can adapt to multiple image formation models as shown in the yellow boxes of Figure 1. For example, the prevalent approach taken to remove artifacts that arise from an unseen image formation model at test time is to re-synthesize the training dataset with the appropriate degradation operator $A_{test}$ and re-train the DNN accordingly (see for example [3], [7]). Other approaches include the development of new training strategies that can adapt to multiple degradation levels [8]. Unfortunately, these approaches possess the significant drawback of being excessively time-consuming and requiring large computational resources. Modifying the training procedures (such as gathering new training data or implementing new loss functions) of neural networks may take weeks, sometimes months, of research to converge to a satisfying solution. In many settings, such as industry, this approach would not be an effective use of time. This also requires access to large computing power and training datasets, which a user may not necessarily have access to at test time. In addition, these approaches are not always guaranteed to succeed as training with multiple degradation operators A does not include all possible operators which may be later encountered at test time. Moreover, increasing the number of A operators during training may degrade the individual performance of the network on each A, as specialization is not possible.

Recently, internal learning methods have been proposed in which no offline training is required as the SR task at test time is entirely learned from a randomly initialized pre-determined CNN [6], [9]. These approaches bring us one step closer to combining analytical SR with DL-based SR as they use priors naturally encoded in CNNs to learn an SR solution that is consistent with the observed data. Furthermore, these frameworks do not suffer from lack of generalization issues as the solution is fully re-computed and specific to each observed LR at test time. However, one drawback of such methods is that the powerful ability of CNNs to learn complex SR functions from large training datasets and deep perceptual loss functions is essentially ignored in their proposed framework. In this paper, we propose a fine-tuning algorithm that combines the powerful learning abilities of deep SR CNNs with the rigor obtained by employing internal learning from the image formation

Preliminary results of this work were presented at the 2019 IEEE International Conference on Image Processing (ICIP) [1]. This work was supported in part by the Sony 2016 Research Award Program Research Project and by the National Science Foundation under grant DGE-1450006. The work of SLT and RM was supported by the Spanish Ministry of Economy and Competitiveness through project DPI2016-77869-C2-2-R and the Visiting Scholar program at the University of Granada. SLT received financial support through the Spanish FPU program. A. Lucas and A.K. Katsaggelos are with the Dept. of Electrical and Computer Engineering, Northwestern University, Evanston, IL, USA. S. Lopez-Tapia and R. Molina are with the Computer Science and Artificial Intelligence Department, Universidad de Granada, Spain.

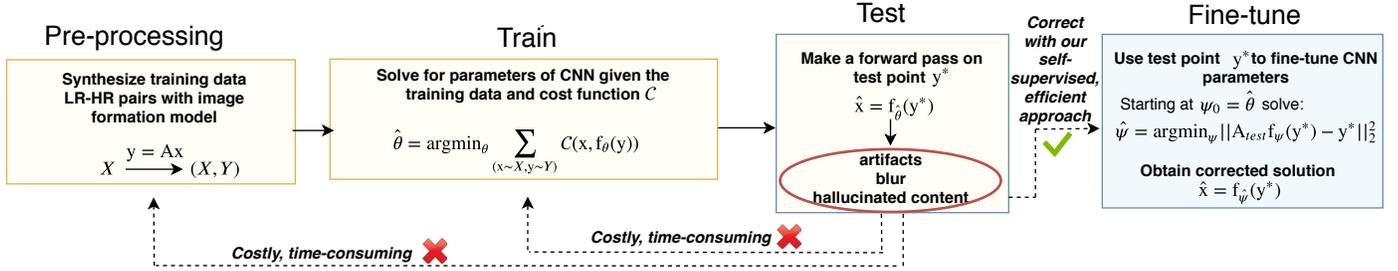

**Figure 1:** An illustration of our approach. We address the case when the image predicted by the CNN contains artifacts (Section IV-A), blur (Section IV-B) or hallucinated content (Section V) . We fine-tune the parameters of the CNN using the observed data point only and avoid the cost of re-training the CNN.

model. Our fine-tuning algorithm addresses the limitations that may arise from learning-based methods and corrects for these using analytical knowledge of the SR problem at test time.

Assuming access to the parameters $\theta$ of a trained SR CNN, we show that one can effortlessly guide these parameters to another set of parameters $\psi$ which outputs a more appropriate solution *without training the CNN on a new dataset*. In our first set of experiments, we show that by letting a trained CNN "overfit" the low-resolution image seen at test time and its corresponding image formation model, one can naturally converge to a perceptually pleasing SR solution. We show that with our proposed approach, it is possible to use an SR CNN to super-resolve an image at a different scale factor than the one it was originally trained for (Section IV-A). Similarly, we employ our method to adapt an SR CNN trained for super-resolving bicubically downsampled LR images to successfully super-resolve ones that have undergone blur and downsampling, both in a blind and non-blind settings (Section IV-B). In the second set of experiments presented in our paper (Section V), we show that our proposed framework is not restricted to solving domain adaptation tasks, but can also correct for various types of artifacts that arise from SR CNNs trained with generative adversarial networks (GANs) [10]. Indeed, while the application of GANs to inverse imaging problems has resulted in solutions of previously unseen restoration quality, including SR tasks ([11], [2], [12], [5]), GANs are also known to frequently generate unpleasant content in their output. For example, it was shown that the popular EnhanceNet (ENet) model may, in some cases, hallucinate content that is not in agreement with the ground-truth HR image [13]. Similarly, the VSRGAN model introduced by [11] generates video frames of high perceptual quality, with the unfortunate simultaneous generation of a high-frequency dot pattern. While these GAN-related artifacts vary in their cause and nature, we show in Section V that our self-supervised fine-tuning framework can naturally correct for these.

## II. Related works

Our work establishes a general-purpose framework for correcting the solution learned by any SR CNN model trained on a large dataset, with the generic objective of correcting artifacts generated by the CNN, regardless of the cause, or nature, of these artifacts. As a result, our method can be applied to a diverse range of applications, including domain adaptation or removal of artifacts originating from GANs. To the best of our knowledge, there exists no publication that has worked with this particular objective. Indeed, while most of the current SR literature focuses on achieving a new state-of-the-art by introducing yet another SR CNN, what we propose here is a generic *tool* to correct existing and SR models in today's SR literature and avoid lengthy training procedures.

Our proposed framework employs an objective function that fully relies on the information available in the low-resolution image and its image formation model at test time. Objective functions of such self-supervised nature have already been introduced in today's SR literature, mainly under two different forms. The first type of internal training in today's literature learns from the internal data that are present in the observed LR image by extracting multi-scale self-examples in the test LR. For example to improve the performance of their SR CNN, [14] uses this function as part of a fine-tuning step. In their work on zero-shot super-resolution (ZSSR), [6] create a "training" dataset adapted to the test data by successively downsampling the LR observation. The resulting pairs of higher and lower resolutions versions of the LR image are then used as sample pairs for training their neural network. Similarly, [15] fine-tune their denoiser CNN on clean and noisy patches extracted from the input image at test time which is then applied in their Plug-and-Play model [16] for image SR.

Another type of internal learning seen in today's literature consists of designing an objective function which utilizes the data fidelity term $Af_\theta(y) = y$ and making use of the information encoded in the forward model of super-resolution: $y = Ax$. The popular Deep Image Prior (DIP) [9] method employs this data consistency constraint to define the minimization problem $\hat{\theta} = \arg\min_{\hat{\theta}} ||Ah_{\hat{\theta}}(z) - y||^2$ using a randomly initialized U-Net $h_\theta(\cdot)$, a fixed random input vector $z$, and the known degradation operator $A$. Through the use of gradient descent, the algorithm eventually converges to a satisfactory restored image: $\hat{x} = h_{\hat{\theta}}(z)$. The supervision used during "training" is solely given by the data consistency constraint $Ah_{\hat{\theta}}(z) = y$. The authors of DIP claim that the architecture of $h_{\hat{\theta}}(\cdot)$ is a strong enough regularizer to produce pleasing images without necessitating further regularization. Indeed, the obtained solutions in [9] look surprisingly similar to natural images. In their work on image-adaptive GANs (IAGAN) for super-resolution and compressed sensing (CS) of face images, the authors of [17] apply such a data-fidelity term to fine-tune and correct for their GAN solution at test time. They show that fine-tuning powerful face-generating GANs on the data fidelity constraint leads to strong restoration performance for super-resolving degraded face images.

Our work has overlap with ZSSR [6] and DIP [9] in that our objective function used fully relies on the internal data available at test time. With this approach, we are not restricted to SR settings assumed during training of a CNN. However, our work utilizes this objective function within a *fine-tuning* framework, which enables us to start from the SR solution provided by pre-trained SR CNN as opposed to learning the SR solution from random parameters as in ZSSR and DIP. Thus our method combines the learning power of pre-trained DNNs with the dependability of analytical approaches for image restoration. With this framework, one can correct already learned SR solutions to new SR settings previously unseen at test time using internal learning. As our approach is not tied to a specific SR model or architecture, it is meant to be used as a "wrapper" function that adapts the solution of any CNN to the new desired objective at test time.

To some extent, our work is related to [17]'s work on IAGANs, as they propose a similar fine-tuning approach to obtain improved SR and CS results. However, because their fine-tuning framework heavily relies on access to a strong GAN prior, their method is currently only applicable to super-resolving faces. Indeed, generalizing their framework to natural image super-resolution would require access to a strong GAN prior for high-resolution natural images, which is not currently feasible. By contrast, the effectiveness of our method is not dependent on the use of such GAN priors.

To summarize, our work differs from the other works in today's literature in that we provide a generic framework to solve a multitude of problems and limitations that may arise when training SR CNNs. It can be used to correct a trained SR CNN at test time regardless of the architecture or objective functions employed during training. Ultimately, the objective of our work is to introduce a simple, efficient tool to today's DL scientists and engineers who wish to enhance the solution to a particular input without performing any amount of training.

## III. METHOD

### A. Self-supervised fine-tuning for correcting SR CNNs

Suppose access to a super-resolving CNN $f_\theta(\cdot)$ trained on a large dataset $(X, Y)$ with objective function $\mathscr{C}$. The datasets used to train these networks are usually generated with the specification of one (or multiple) image formation models, $y = Ax$. Typically in super-resolution, the degradation operator A corresponds to a bicubic down-sampling operator for a specific scale factor. As a result of training, we obtain the estimated parameters $\hat{\theta}$ by solving $\hat{\theta} = \arg\min_\theta \mathbb{E}_{X,Y} \mathscr{C}(x, y, \theta)$ where the high and low-resolution pairs $x \sim X$ and $y \sim Y$ are sampled from the training dataset. In the rest of our paper, we do not constrain $f_\theta$ to correspond to a specific CNN architecture, nor do we assume a particular training criterion $\mathscr{C}(x, y, \theta)$.

In the experiments described below, we wish to obtain a new set of parameters $\psi$ from the original set of pre-trained parameters $\theta$ that adapts to a different SR objective at test time. In Section IV for example, the objective at test time is to adapt the available SR CNN trained on a specific image formation model to another image formation model (e.g., super-resolving for a different scale factor in Section IV-A). Inspired by the work in [9], we argue that the information required by the SR CNN to adapt to the new SR problem at test time can be extracted from the formulation of the image formation model at test time, $y = A_{test}x$. Thus we can encourage the SR CNN to satisfy this requirement and obtain an improved set of parameters for the problem at test time with the following data-fidelity fine-tuning objective function:

$$\hat{\psi} = \arg\min_\psi ||A_{test}f_\psi(y) - y||_2^2, \quad (1)$$

where the starting vector $\psi_0$ for this optimization problem is set to $\hat{\theta}$, and $A_{test}$ is the degradation operator at test time (it differs from the A used to generate the training dataset). The pseudocode for solving Eq. 1 is shown in Algorithm 1. In Section V, we address the problem of removing artifacts hallucinated by GANs for super-resolution problems. As these artifacts are not consistent with the constraints provided by the image formation model used at train time, we hypothesize that we can naturally remove these artifacts by enforcing the SR GAN to overfit the data-fidelity term introduced earlier:

$$\hat{\psi} = \arg\min_\psi ||Af_\psi(y) - y||_2^2, \quad (2)$$

where in this setting $A = A_{test}$ corresponds to the degradation operator assumed during training. We note here that our formulation in Equations 1 and 2 corresponds to the well established inversion problem posed by all image recovery tasks, applied here to the context of fine-tuning and output enhancement of an already trained network.

### B. Optimization details

We use the Pytorch [18] Deep Learning library for performing the experiments detailed below. As seen in Algorithm 1, we automatically stop our optimization using a stopping criterion. We compute the "low-resolution" PSNR metric on the predicted low-resolution $A_{test}f_\psi(y)$ and the ground-truth observed y and employ it as our stopping criterion for the experiments in Section IV. More specifically, we automatically interrupt the optimization when the PSNR on the LR data does not improve by more than $\delta = 0.04$ dB.

Our stopping criterion for the GAN experiments described in Section V corresponds to the Perceptual Similarity (PS) metric (introduced below in Section III-D), again computed with respect to the low-resolution information. In this case, we interrupt our fine-tuning framework when the PS metric does not decrease by more than $\delta = 3 \times 10^{-5}$. These values were determined experimentally. We set the maximum number of iterations to $K = 4000$.

We set the learning rate to $\alpha = 0.01$ for all of our experiments and use the Stochastic Gradient Descent (SGD) optimizer in PyTorch [18] with momentum $\mu = 0.9$. No additional regularization is used. Note that computing the gradients of the terms in Equations 1 and 2 necessary to perform gradient descent is straightforward with the automatic differentiation functionalities offered by Pytorch and other available Deep Learning libraries. With the converged fine-tuned parameters $\hat{\psi}$, we then use $\hat{x} = f_{\hat{\psi}}(y)$ as the final enhanced, artifact-free solution.

**Algorithm 1** Our proposed fine-tuning algorithm.

1: **function** FINETUNE($\hat{\theta}$, y, $A_{test}$)
2:     $\psi \leftarrow \hat{\theta}$                         ▷ Initialize parameters
3:     *earlyStop* ← **False**    ▷ Initialize stopping condition
4:     **while** *earlyStop* **False do**
5:        $x \leftarrow f_\psi(y)$                        ▷ Forward pass
6:        $\mathcal{L} \leftarrow ||A_{test}x - y||_2^2$              ▷ Eq. 1
7:        Compute $\nabla_\psi \mathcal{L}$               ▷ Backward pass
8:        $\psi \leftarrow \psi - \alpha \nabla_\psi \mathcal{L}$         ▷ Update weight
9:        update *earlyStop*        ▷ Check convergence
       **return** $f_\psi(y)$.               ▷ Compute final SR

### C. Models

Let us re-iterate here that our proposed approach is not tied to a specific SR architecture but instead can *a priori* be applied to any existing SR models. Below we apply our fine-tuning framework on multiple CNNs $f_\theta(\cdot)$ introduced in today's DL literature, which all differ in their architectures and training loss functions. We briefly describe these models below.

Our experiments in Section IV employ the EDSR [3] network, one of today's state-of-the-art image super-resolution network. The EDSR network consists of several deep residual layers followed by learnable upsampling modules at the end of the model that output a final SR image of the desired HR spatial extent. The EDSR model was trained on the Mean-Absolute-Error (MAE) function $\mathcal{C}_{MAE} = ||f_\theta(y) - x||$. For our experiments we download the EDSR model trained for scale factor 4 from the official Pytorch GitHub repository https://github.com/thstkdgus35/EDSR-PyTorch.

In addition to the EDSR model, we apply our framework on the EnhanceNet (ENet) [13] model. Similarly to EDSR, it is composed of a series of residual layers, however, fixed interpolation layers are used to increase the spatial extent of the input as opposed to learnable upsampling modules. During training, the ENet employs a combination of texture $\mathcal{C}_{text} = ||G(\rho(f_\theta(y))) - G(\rho(x))||_2^2$, feature $\mathcal{C}_{feat} = ||\rho(f_\theta(y)) - \rho(x)||_2^2$, and adversarial losses $\mathcal{C}_{GAN} = \log(1 - g_\phi(f_\theta(y)))$ where $\rho(\cdot)$ is a pre-determined feature-space obtained from a pre-trained discriminative CNN and $g_\phi(\cdot)$ is a discriminator network with trainable parameters $\phi$. We download the ENet model trained for scale factor 4 from the official Tensorflow repository https://github.com/msmsajjadi/EnhanceNet-Code and convert the resulting Tensorflow model to Pytorch.

Finally, we test our framework on our own video SR methods. More specifically we obtain our pre-trained neural networks from [11] which introduces two models trained for the task of VSR for scale factor 4. In this paper, we refer to these two models as the VSRMSE model, trained with the Mean-Squared-Error loss $\mathcal{C}_{MSE} = ||f_\theta(y) - x||_2^2$, and the VSRGAN model trained with a combination of feature and adversarial loss similar to ENet's objective functions. Both the VSRMSE and VSRGAN architectures are composed of 15 residual blocks (See Figure 1 in [11]). Because in this work we employ our video SR networks in a still-image setting as opposed to videos, the input to our networks corresponds to replicates of the bicubically-interpolated LR image y = {y,y,y,y,y} and the output is the corresponding SR estimate x.

We conclude this section with a remark regarding the notation used in the rest of our paper. We will refer to an SR CNN fine-tuned with Equation 1 or 2 by juxtaposing an asterisk ∗ with the name of the model. For example, EDSR refers to the original model $f_\theta(\cdot)$ originally trained and made available by the authors [3], whereas EDSR∗ denotes the model $f_\psi(\cdot)$ obtained as a result of applying our fine-tuning framework.

### D. Metrics and datasets for quantitative assessment

To quantitatively assess the performance of our self-supervised fine-tuning framework, we compute two metrics and report these below each figure and table presented in this paper. The first metric we report is the well-known peak signal-to-noise ratio (PSNR) [19] with respect to the high-resolution target image. While the PSNR metric is the de facto standard metrics employed for assessing the performance of an image restoration model, it does not always provide an accurate assessment of the perceptual quality of images produced by deep neural networks. Therefore, in addition to the PSNR, we also provide the Perceptual Similarity (PS) metric proposed by [20]. This metric leverages the representations learned by CNNs to output a distance value between a reference and ground-truth image. The smaller the distance, the better the restoration quality. For more information regarding this metric, we refer the reader to [11], [21]. We remark here that while the PS metric's popularity is increasing in the SR literature ([11], [22], [17]), we did not find it to be *consistently* accurate in comparing the restoration quality of two images. Thus as any metric, the PS is to be used in addition to a careful qualitative visual examination of the results. For all the experiments considered in our paper, we report these two metrics on three widely used benchmark datasets, namely the Set5 [23], Set14 [24] and BSD100 [25] datasets.

## IV. Correcting for disagreement in image formation models

We first consider the case in which the image formation model at test time differs from the one assumed during training, i.e. $y = A_{test}x$ and $A_{test} \neq A$. In Section IV-A, we show that we can adapt CNNs trained for super-resolving an LR input by factor 4 to learn to super-resolve an LR by scale factor 2 or 3 by self-correcting it with equation 1. Similarly in Section IV-B, we show that we can fine-tune a CNN to super-resolve an LR that underwent a blur degradation even if the CNN was never trained to account for blur at its input.

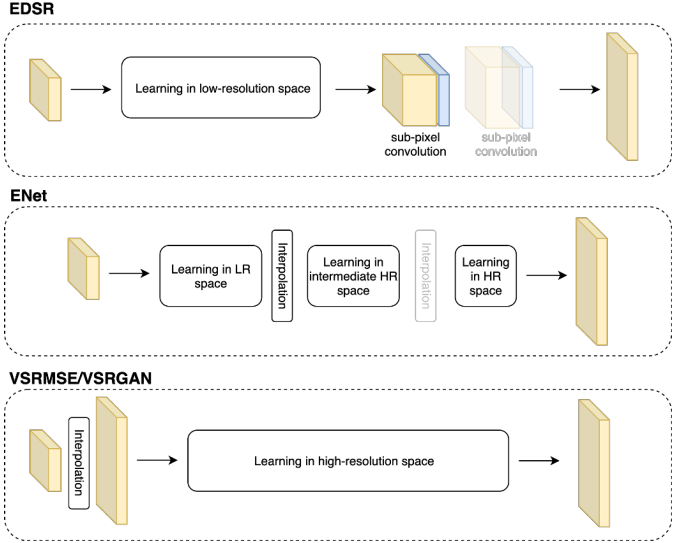

**Figure 2:** Adapting SR CNN architectures designed for scale factor 4 to super-resolve by a scale factor 2. In EDSR, the second upsampling sub-pixel convolution module is removed. In ENet, the second interpolation layer is removed. In VSRMSE/VSRGAN, no modification is necessary as the LR is bicubically interpolated to the correct spatial extent at prior to being inputted to the network.

### A. Disagreement in scale factors

Let us first consider the case in which we would like to correct an SR CNN network trained for factor 4 to perform the SR problem of scale factor 2. In this section, we focus on three networks: the EDSR [3], ENet [13], and VSRMSE [11] networks all trained for scale factor 4, i.e., in this case the A used to generate their respective training datasets corresponds to the bicubic downsampling operator for scale factor 4. We first note that because the EDSR and ENet contain layers in their architecture that increase the spatial extent of the input by 4, such scale-specific layers need to be modified to have these networks produce an output whose final shape is consistent with the new desired scale factor of 2. A visualization of how we modify the EDSR and ENet architectures to adapt to the scale factor 2 SR problem is shown in Figure 2.

The top row of Figure 3 shows the effect of trying to super-resolve an LR image by a factor of 2 using EDSR, ENet or VSRMSE, each trained for scale factor 4. It is clear from this figure that while the model is successful at increasing the resolution of the low-resolution image, artifacts appear as a result of super-resolving for a scale factor for which the network was not trained. It is safe to say that these artifacts make the resulting output an unsuitable SR solution. What is particularly interesting is that these artifacts vastly differ in their nature, from model to model. The EDSR network's "artifacts" are not artifacts *per se* as we the resulting color space is wrong, however, the perceptual quality of the SR image is surprisingly good. We hypothesize that this change in color space is the result of removing the second upsampling module (as shown in the first row of Figure 2) to adapt the EDSR network to the new scale factor. This upsampling module contains a learnable convolution operation that may be responsible for mapping to the correct color space. In addition, we observe that the VSRMSE artifacts in Figure 3(c) are more severe than the ones observed in ENet in 3(b). This effect is caused by VSRMSE operating in the high-resolution space as opposed to low-resolution space: as there is a mismatch in sale factors at the input layer, the error propagates through the rest of the VSRMSE to result in particularly un-natural images. This hypothesis is consistent with our observation that smaller amount of error is observed in ENet's output, as its first convolution layers operate in low-resolution space before being interpolated to the high-resolution space, as shown in Figure 2.

Similar artifacts are seen when inputting an LR image corresponding to scale factor 3 to ENet and VSRMSE trained for scale factor 4, as seen in the bottom row of Figure 3(b) and 3(c). Note that modifying the EDSR architecture to perform the SR problem of scale factor 3 requires us to replace the two pre-trained sub-pixel convolution modules for scale factor 2 by an untrained upsampling module for scale factor 3. The incorporation of this random convolution layer introduces major artifacts in EDSR's SR output as shown in Figure 3(a). Our fine-tuning approach will not perform well in such cases and thus in our following experiments we do not fine-tune EDSR for scale factor 3. The ENet and VSRMSE, however, do not have this limitation as as they do not contain learnable upsampling layers that are specific to the scale at test time.

In the cases in which artifacts are not severe, we hypothesize that some of the knowledge learned by EDSR, ENet, and VSRMSE networks trained for a factor of 4 may still be useful for super-resolving LR images for factors 2 or 3. Thus our current objective is to enhance the perceptual quality of these output images while keeping the observed increase in resolution seen in Figure 3. In other words, our objective is to obtain a corrected SR solution specific to the new scale factor at test time for this particular input LR image using self-supervised fine-tuning. While the artifacts seen in EDSR, ENet, and VSRMSE are vastly different in nature, we use the *same* self-supervised fine-tuning method defined in Equation 1, with the same objective function and same optimization hyper-parameters to obtain an SR solution corrected for the new scale factor. In the fine-tuning formulation introduced in Equation 1, y is the LR image downsampled by scale factor 3, $A_{test}$ is the bicubic downsampling operator for the new scale factor at test time (2 or 3) and the f(·) network is the EDSR, ENet, or VSRMSE network trained for scale factor 4. We set the parameters $\psi_0$ at the first iteration to correspond to the pre-trained $\hat{\theta}$ parameters of each of the three networks. As a result of fine-tuning, we are provided with our new parameters $\hat{\psi}$ and thus generate a new solution $\hat{x} = f_{\hat{\psi}}(y)$ denoted as EDSR∗, ENet∗ and VSRMSE∗ and displayed in the second columns of Figures 4, 5 and 6. Below each of these figures we report the PSNR and PS metrics computed on the corresponding test image. Figure 4 shows that the SR image produced by EDSR is mapped pack to the correct color space with the fine-tuned EDSR∗. Figures 5 and 6 reveal a considerably less amount of distortions in their super-resolved solutions of ENet∗ and VSRMSE∗, respectively. These figures, along with the metrics reported below each of them, show that the artifacts originally observed are successfully removed. In other words, in all three cases, we obtain a perceptually pleasing SR solution for scale factor 2, while none of these networks were trained to do so.

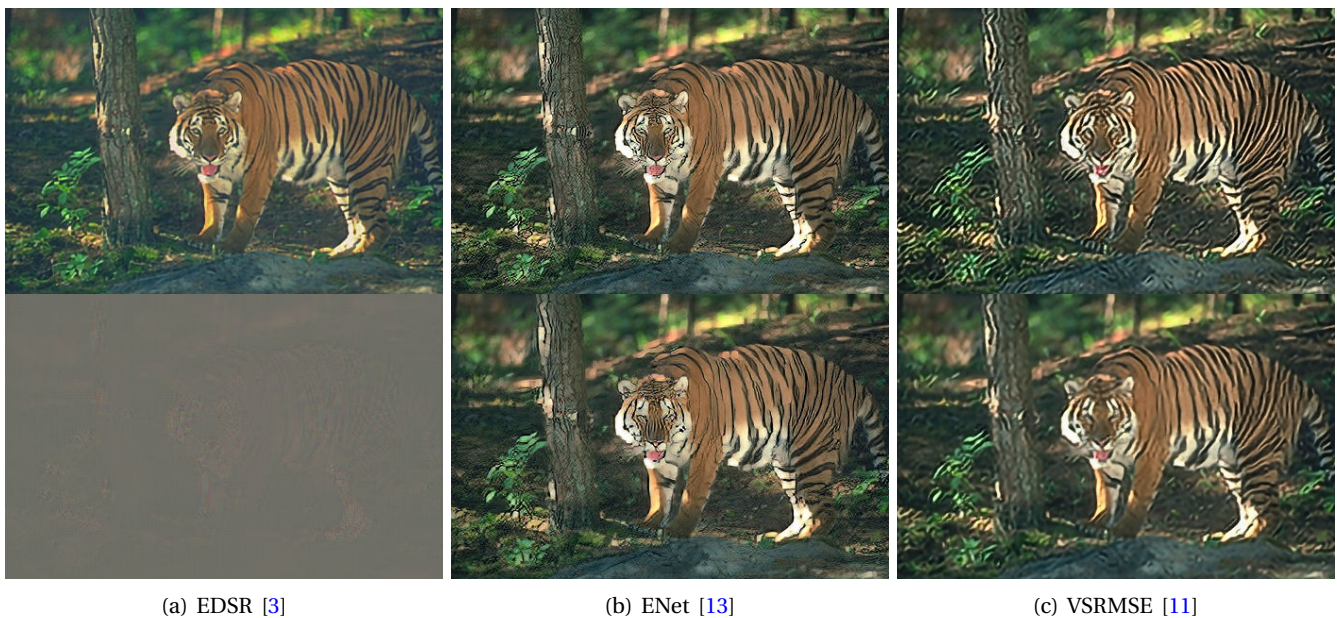

(a) EDSR [3]  (b) ENet [13]  (c) VSRMSE [11]

**Figure 3:** Outputs of SR models when trained for scale factor 4 and super-resolving for factor 2 (top row) and 3 (bottom row). The poor solution obtained by EDSR is the consequence of replacing a trained upsampling module with a randomly initialized upsampling module for adapting to the scale factor 3 problem (See Section IV-A for more detail).

|  | | **Bicubic** | | **VSRMSE∗** | | | | **ENet∗** | | | | **EDSR∗** | | | |
|---|---|---|---|---|---|---|---|---|---|---|---|---|---|---|---|
|  | | | | PSNR | | PS | | PSNR | | PS | | PSNR | | PS | |
|  | Desired Scale | PSNR | PS | Start | End | Start | End | Start | End | Start | End | Start | End | Start | End |
| Set5 | 2 | 33.69 | 0.0479 | 24.61 | 34.68 | 0.1005 | 0.0337 | 29.65 | 35.49 | 0.0208 | 0.0116 | 21.72 | 36.92 | 0.0307 | 0.0137 |
|  | 3 | 30.44 | 0.1021 | 28.83 | 32.56 | 0.0627 | 0.0507 | 28.62 | 31.48 | 0.0299 | 0.0206 | N/A | N/A | N/A | N/A |
| Set14 | 2 | 30.35 | 0.0676 | 22.82 | 31.09 | 0.1284 | 0.0556 | 25.91 | 31.82 | 0.0453 | 0.0279 | 21.67 | 33.00 | 0.0463 | 0.0288 |
|  | 3 | 27.63 | 0.1370 | 26.35 | 29.12 | 0.0989 | 0.0902 | 25.11 | 28.38 | 0.0584 | 0.0500 | N/A | N/A | N/A | N/A |
| BSD100 | 2 | 29.49 | 0.0961 | 22.90 | 30.01 | 0.1653 | 0.0842 | 24.18 | 30.64 | 0.0624 | 0.0484 | 22.41 | 31.69 | 0.0616 | 0.0486 |
|  | 3 | 27.14 | 0.1754 | 26.28 | 28.07 | 0.1356 | 0.1313 | 24.22 | 27.54 | 0.0695 | 0.0746 | N/A | N/A | N/A | N/A |

TABLE I: Average PSNR and Perceptual Similarity (PS) before and after fine-tuning VSRMSE, ENet, and EDSR for each dataset. The experiments are described in Section IV-A. We do not report the results of fine-tuning EDSR for scale factor 3 as the original solution $f_\theta(y)$ presents severe artifacts that are not appropriate for fine-tuning (Section VI-E addresses such limitations in more detail).

We next fine-tune EDSR, ENet and VSRMSE on all images of the Set5, Set14 and BSD100 datasets described in Section III-D for learning the SR problems of scale factors 2 and 3 on each LR image. We report the PSNR and PS metrics corresponding to the original $f_{\hat{\theta}}(y)$ (at the start of fine-tuning) solution and those corresponding to the fine-tuned $f_{\hat{\psi}}(y)$ (when our fine-tuning algorithm has converged). In Table I we report the average of these computed metrics over all images in the datasets. The metrics in Table I show a consistent quantitative improvement as a result of fine-tuning, revealing a sharp increase in PSNR and a significant decrease in PS across all datasets and models. Furthermore, we show in Table I the average PSNR and PS metrics computed on the bicubically interpolated low-

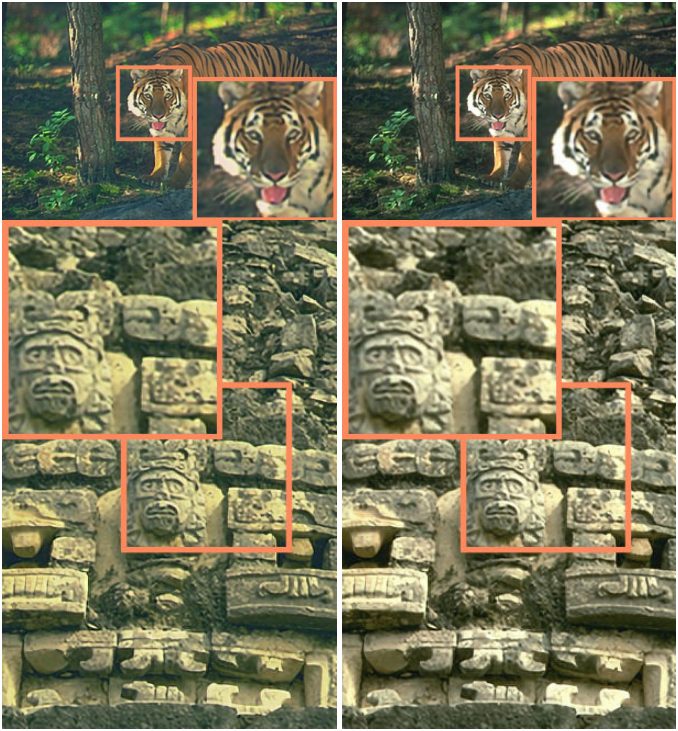

(a) EDSR (24.46/0.0493)    (b) EDSR∗ (31.83/0.0355)

**Figure 4:** Correcting EDSR for scale factor 2 (Section IV-A).

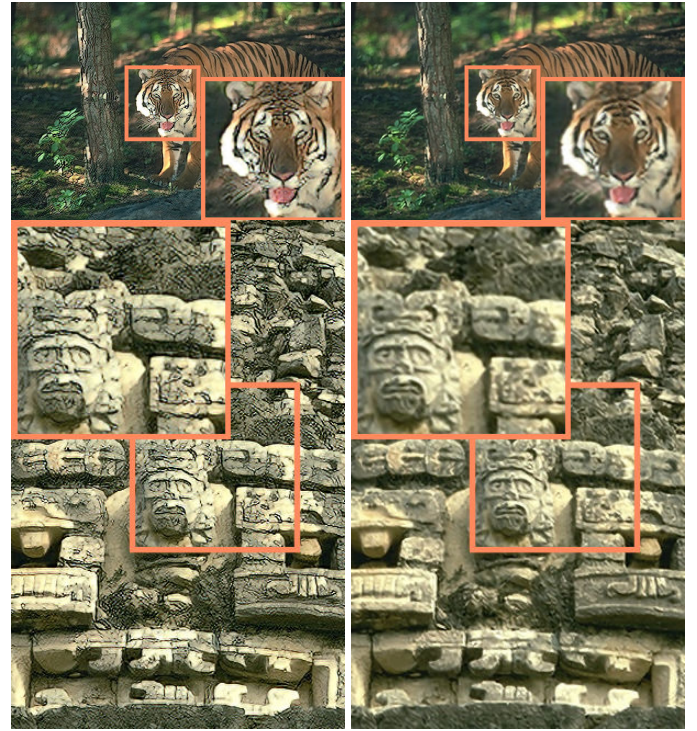

(a) ENet (23.83/0.0437)    (b) ENet∗ (30.72/0.0355)

**Figure 5:** Correcting ENet for scale factor 2 (Section IV-A).

resolution images. Comparing these results with the quantitative results obtained from EDSR∗, ENet∗, and VSRMSE∗ at convergence confirms that an SR solution for scale factors 2 and 3 was successfully learned in all cases.

*B. Disagreement in blur operator*

The previous section assumed a forward process in which the high-resolution image is downsampled with bicubic downsampling D, i.e., y = Ax = Dx. This coincides with today's prevailing approach in DL to generating pairs of high and low-resolution images with such a formation model. A more general and accurate model for SR is one in which blur is applied prior to downsampling, i.e. y = Ax = DBx. Indeed low-resolution images may also suffer from blur for example due to motion, optics, sensor, or atmospheric turbulence [26]. In these cases, super-resolving such images with a CNN trained without taking such possible blur deformations into account will result in an SR image with excessive blur. For the following set of experiments, we focus on the EDSR [3] network trained for a scale factor of 4 which was never trained to remove blur. Our objective is to correct the EDSR solution at test time to super-resolve its input even when blur is present. Our experiment below considers both the blind (in which case we do not know the blur B) and non-blind (in which case we know the blur B) setting.

On their work on zero-shot super-resolution (ZSSR), [6] generate what they refer to as a "non-ideal" LR dataset by creating random downsampling Gaussian kernels of various shapes and sizes and applying these on HR images in the BSD100 dataset (see Section 4.2 in their paper for details). We follow their generation process and generate Gaussian blur kernels in a random fashion and apply each resulting kernel to a corresponding image in the BSD100 dataset to obtain our non-ideal LR dataset. It is clear that because EDSR was not trained to take such complex blurs into account, it will fail at properly super-resolving the input LR. An example of such an EDSR output is shown in the first column of Figure 7. As expected, the resulting images are particularly smooth and lack detail. The blurry LR images were not successfully super-resolved by EDSR.

Typically the approach taken to obtain an SR CNN capable of removing complex blurs from the input LR is to either re-synthesize an appropriate training dataset such that the SR CNN can be trained on the new domain, or revert to using an internal learning approach such as ZSSR [6]. We propose an alternative approach in which we fine-tune the original EDSR network on the data fidelity loss in Equation 1 where we define $A_{test}$ = DB where B is the Gaussian blur generated for the test image $y_{test}$ and D is bicubic downsampling. As a result, the new EDSR network learns a corrected SR function that performs joint deblurring and super-resolution for this particular test image and image formation model.

Let us first consider the non-blind setting, in which we assume that the blur operator B is known. With our knowledge of B, we apply our proposed fine-tuning framework to obtain EDSR∗ corrected to perform joint deblurring and super-resolution on its input. For comparison, we apply the official ZSSR code made available by the authors at https://github.com/assafshocher/ZSSR on our synthesized

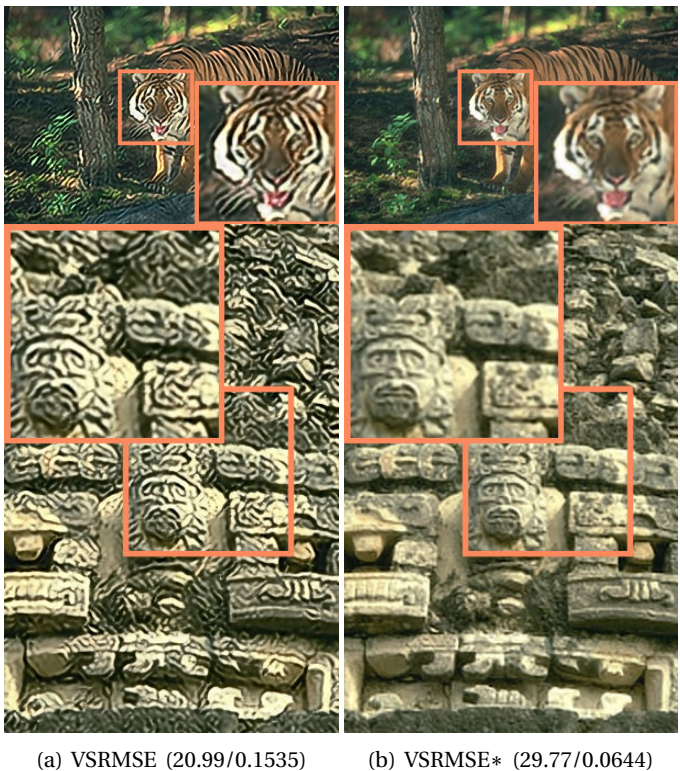

(a) VSRMSE (20.99/0.1535)  (b) VSRMSE∗ (29.77/0.0644)

**Figure 6:** Correcting VSRMSE for scale factor 2 (Section IV-A).

| Setting | EDSR [3] | | ZSSR [6] | | EDSR∗ | |
|---|---|---|---|---|---|---|
| | PSNR | PS | PSNR | PS | PSNR | PS |
| **Non-blind** | 19.83 | 0.2981 | 25.52 | 0.2131 | 25.80 | 0.2388 |
| **Blind** | 19.83 | 0.2981 | 24.88 | 0.2292 | 24.48 | 0.2587 |

TABLE II: Average PSNR and PS as a result of running EDSR, ZSSR, and fine-tuned EDSR∗ described in Section IV-B. The LR dataset is generated by applying Gaussian kernels on various shapes to the HR images in BSD100.

non-ideal LR dataset. We report the average metrics computed on the super-resolution results provided by EDSR, the fine-tuned EDSR∗ and ZSSR in the first row of Table II. A significant increase in PSNR (near 6 dB increase) is observed as a result of fine-tuning EDSR, along with an improvement in the PS. We note that in this setting EDSR∗ and ZSSR show similar quantitative performance.

However, it is often the case that the image formation model, and more specifically the Gaussian blur B in this experiment, is not known. In such cases, the unknown blur may be computed using techniques available in the image processing literature [26], [27], [28], [29], [30]. We follow this approach and estimate B from the observed LR images using the algorithm and code provided by [30]. With our estimates of Gaussian kernels for each LR image, we apply our fine-tuning algorithm in Equation 1 on each LR image in the dataset. In order to compare ZSSR and EDSR∗ in the blind setting, we run the ZSSR code using these estimates as well. We report the resulting metrics in the second row of Table II. While the computed PSNR is lower than the ones computed in the non-blind setting, the PS metric does not decrease as much. These observations suggest that our fine-tuning algorithm performs well even with inexact estimates of the blur B.

We provide a qualitative comparison of our self-supervised fine-tuning framework and the ZSSR approach in Figure 7. The metrics computed individually for each images are reported below each figure. The qualitative results displayed in Figure 7 are consistent with the earlier quantitative observations, as EDSR∗ and ZSSR produce results of similar perceptual quality.

## V. CORRECTING GENERATIVE ADVERSARIAL NETWORKS WITH FINE-TUNING

In the previous section, we have shown that one can use a knowledge (or estimate) of the image-formation model to adapt an SR CNN to a new SR problem unseen during training. We now depart from addressing the problem of disagreements in image formation model. Instead, we consider the case in which a produced SR result at test time is of unsatisfactory quality, either due to artifacts generated or due to inaccurate content when compared with respect to the ground-truth HR content. More specifically, we look into SR images computed by Generative Adversarial Networks (GANs), which have become an increasingly popular tool for super-resolving images and videos. Often combined with textural and perceptual losses [13], [31], these adversarially trained networks are trained to accept an LR image as their input and make the SR output appear *as if* it was sampled from the high-resolution data [10], [4]. Despite their success in generating highly realistic images, GANs trained for super-resolution applications often suffer from introduced undesirable artifacts [11], [4]. These artifacts may arise due to the unregularized nature of adversarial loss functions (Section V-A), or may appear due to complex training dynamics (Section V-B). In the section below we show that our our fine-tuning framework defined in Equation 2 can successfully regularize such GAN models post-training.

### A. Correcting for hallucinated content in ENet

We first study the ENet [13] network, an SR CNN trained with a combination of texture, feature, and adversarial losses. While SR images produced by ENet look particularly realistic, they fail at matching the ground truth images on a pixel-wise scale. This limitation is addressed in the original ENet paper [13], where the authors hypothesize that the hallucinated content is a result of learning textures that occur frequently between pairs of LR and HR images. Examples of mistakes made by ENet on three images from the Set5 and Set14 images are shown in Figure 8. Comparing the ground-truth (left column in Figure 8) images with the ENet outputs (middle column in Figure 8) reveals that some of the content recovered by ENet, while looking high-resolution, is not in agreement with the ground-truth content.

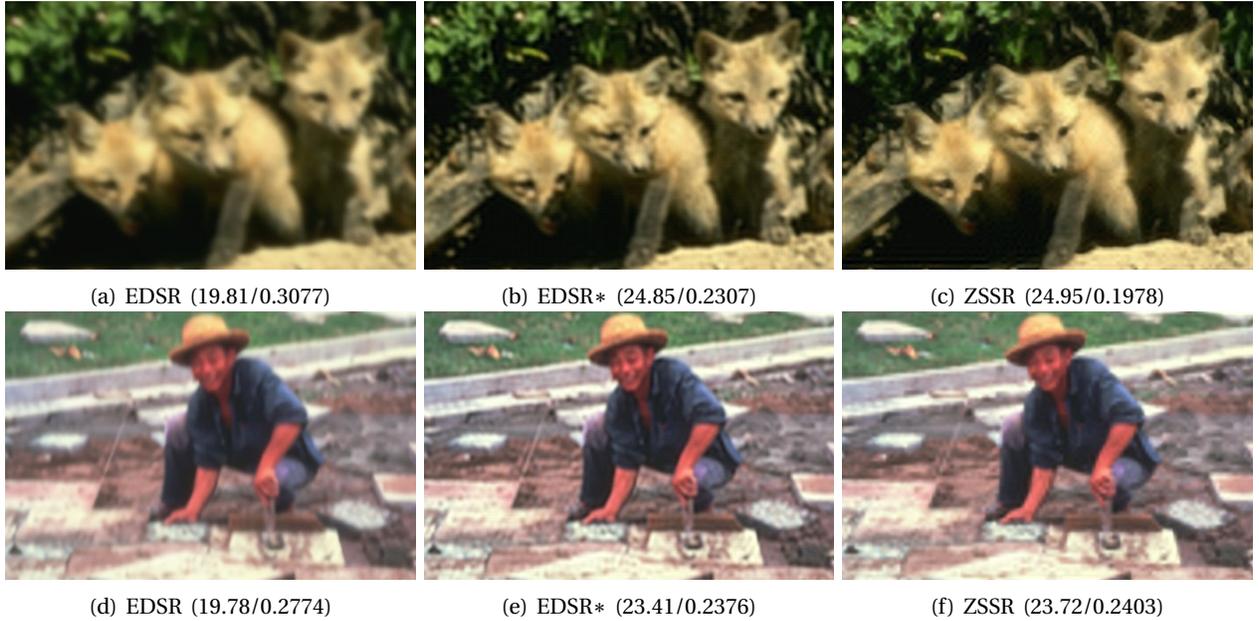

**Figure 7:** (a,d) The original SR obtained by EDSR [3] when provided a blurry LR image at its input as detailed in Section IV-B. (b,e) The SR obtained by our fine-tuned EDSR∗ and (c,f) the SR obtained by ZSSR [6].

An SR image of high perceptual quality but of inaccurate content with respect to the ground-truth HR defeats the original objective of the image super-resolution task. While seemingly harmless for natural image super-resolution, hallucinated content in SR images may become much more of an issue in scientific fields such as medical or forensic applications, as the super-resolved content may be used as part of a critical diagnosis. The introduction of a tool that can naturally translate the original SR image to become more consistent with the ground-truth observation would be very impactful in such communities. We argue that as the observed hallucinated content is not in agreement with the information available in the low-resolution image, using the image formation model as new supervision signal to ENet will naturally guide it towards a solution with accurate content. We thus propose to use our fine-tuning method to correct for such hallucinated content.

We employ our fully self-supervised fine-tuning method as defined by Equation 2 as a means to restore the correct pixel-wise ground-truth content in the SR images produced by ENet. Note that in this case our $A_{test}$ at test time does not differ from the one assumed in training, i.e., $A = A_{test}$. Indeed the purpose of the fine-tuning in this case is to use the information available in the observed LR and its corresponding image formation model to output a new SR solution that is in agreement with the ground-truth content. We show the SR result of our fine-tuned ENet∗ in the third column of Figure 8. Comparing these results with the corresponding ground-truth HR shows that our method is capable of restoring the ground-truth texture that was originally missing from the initial solution. The improvement is also observed quantitatively, as the metrics shown below each image in Figure 8 present a sharp increase in PSNR and a similar decrease in the PS metric.

### B. Correcting for artifacts arising due to challenging training dynamics in VSRGAN

As a second application of reducing artifacts produced by GANs in SR, we now focus on artifacts generated by one of our VSRGAN models originally introduced in [11]. In Section V-B1, we show that VSRGAN's generated artifacts occur as a result of high levels of uncertainty in certain regions of the image. We then show in Section V-B2 that our fine-tuning framework is capable of reducing such behavior and efficiently remove these undesirable artifacts.

*1) Dot-pattern artifacts, a symptom of uncertainty:* Examples of images super-resolved by VSRGAN are shown in the first column of Figure 9. Zooming in these images reveals the presence of a "dot-like" noise pattern in some of the high-detail regions. Viewing the videos super-resolved by VSRGAN revealed a significant amount of flickering of these artifacts across consecutive frames. This severe lack of motion consistency of the generated noise suggests a possible relationship between model uncertainty and the produced artifacts. We thus hypothesize that our trained VSRGAN model and possibly other GAN-based models, suffer from model uncertainty when producing the dot-pattern noise in the super-resolved pixels. Below, we derive a method to visualize such uncertainty.

Discriminative Deep Neural Networks trained for image classification tasks have recently received a lot of attention, particularly since they were found to be prone to adversarial attacks, attacks that tailor small changes in the input image with the objective to fool the DNN classifier [32], [33]. These attacks are created by adding a small perturbation to a normal image which causes the classifier to misclassify the sample, but does not look different to the human eye. These harmful input images are often referred to as adversarial inputs, or adversarial samples, in the literature. To prevent such deceitful classifications, numerous works in the current literature are now focusing on the development of methods that automatically detect adversarial inputs [34], [35]. The authors of [35] found that higher levels of uncertainty in the DNN's output were correlated with the presence of an adversarial sample at its input. Thus the authors propose to use dropout [36], a regularization layer that has been linked with Bayesian uncertainty estimation [37], to quantify the uncertainty of a DNN's prediction for a given input and use the results as a means to detect adversarial attacks.

To visualize VSRGAN's uncertainty, we build on the work in [35] and insert dropout layers in VSRGAN at test time to insert randomness into our model and study the consistency of its prediction across multiple forward passes. We add dropout layers in between each of the residual blocks and set the probability of dropping out corresponding neurons at $p = 0.0005$. We note here that while the VSRGAN architecture does not originally contain dropout layers, no re-training stage is required here. At test time, given an LR input, we perform 50 forward passes with our VSRGAN-dropout model and compute the pixel-wise variance across the 50 predicted frames. According to [35], higher levels of variance in a region of the super-resolved frame suggests lower confidence in the model's prediction. We show such a computed variance map for three different LR inputs in the third column of Figure 9 (for visualization purposes, we show high variance points in black pixels, whereas low variance points are encoded in white). Viewing this variance map reveals that model uncertainty is far from being uniform across the predicted pixels. More particularly, high variance regions seem to correspond to the location of the produced artifacts. This observation reinforces our earlier hypothesis that regions of artifacts are associated with lower confidence levels of VSRGAN.

*2) Applying fine-tuning on VSRGAN:* We now show that we can correct for VSRGAN's learned behavior and naturally reduce the observed dot-pattern using our fine-tuning framework proposed in Equation 2. Although VSRGAN's artifacts do not share the same nature nor cause of ENet's artifacts displayed in Section V-A, we still use the same objective function as the one employed earlier for replacing the hallucinated content in ENet. We show the output of the new VSRGAN∗ in the second column of Figure 9. In these figures it is clear that the dot pattern originally observed is entirely removed as a result of the fine-tuning. This comes with a cost of obtaining a smoother image, which is a natural effect of using the mean-squared-error loss in Equation 2 and is reflected in the increased PS metric.

Furthermore, we show in the fourth column of Figure 9 the variance map computed on the fine-tuned VRSGAN∗ output. When comparing the uncertainty map associated with VSRGAN, it is clear that regions of previously high variance (where black pixels clustered) became significantly less dense as a result of fine-tuning. Overall, the variance map computed from VSRGAN∗ appears lighter and more uniformly distributed. We conclude that fine-tuning VSRGAN on the observation model at test time helps VSRGAN toward predicting pixel values with greater accuracy and overall produce cleaner results.

Finally, we apply our fine-tuning algorithm for reducing ENet and VSRGAN artifacts on all images in the Set5 and Set14 datasets and compute the average metrics before and after running the optimization. The quantitative results shown in Table III reveal a consistent improvement in the PSNR metric, which is in agreement with our earlier observation that a more accurate pixel-wise match with the ground-truth is obtained. However, note that there is an increase in PS from VSRGAN to VSRGAN∗. This is a result of the PS metric sometimes favoring undesirable artifacts over images over smoother images. Ultimately, whether artifacts or smooth images should be favored in SR is somewhat subjective and thus we leave the final decision to the reader.

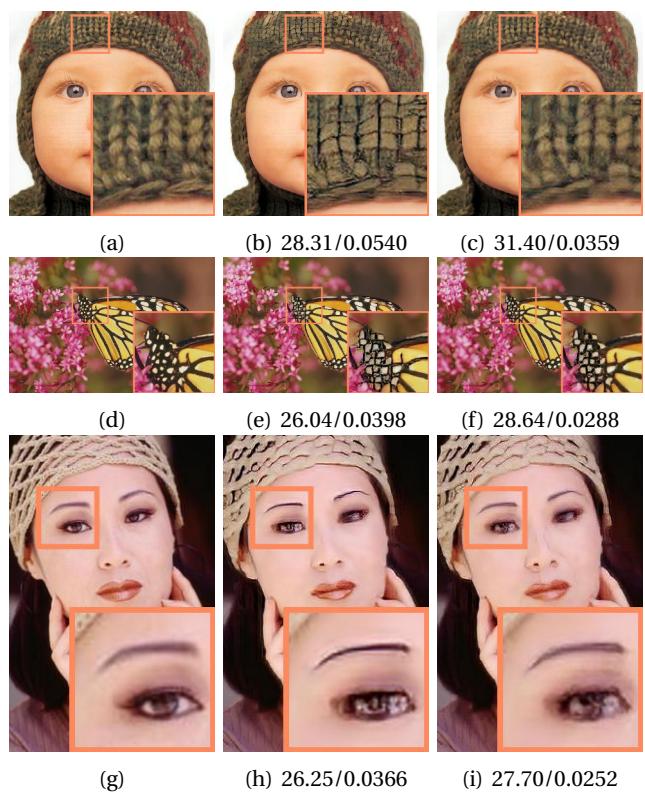

(a)    (b) 28.31/0.0540    (c) 31.40/0.0359

(d)    (e) 26.04/0.0398    (f) 28.64/0.0288

(g)    (h) 26.25/0.0366    (i) 27.70/0.0252

**Figure 8:** Correcting hallucinated content in ENet (Section V-A). (a,d,g) Ground-truth HR; (b,e,h) ENet [13] output; (c,f,i) ENet∗ output.

## VI. DISCUSSION

### A. Advantages

The experimental results presented in Sections IV and V demonstrated that our proposed fine-tuning algorithm is capable of removing artifacts for various applications, e.g., to correct it to a new SR setting (Sections IV-A, IV-B) or to restore accurate content in ENet (Section V-A) and VSRGAN (Section V-B ). The fact that the PSNR and PS metrics simultaneously improved in most cases confirms our algorithm's two-folds ability in increasing (1) the result's fidelity to the ground truth observation and (2) the photorealistic quality of the test image.

|  | Bicubic | | ENet [13] | | | | VSRGAN [11] | | | |
|  | | | PSNR | | PS | | PSNR | | PS | |
|  | PSNR | PS | Beg. | End. | Beg. | End. | Beg. | End. | Beg. | End. |
| --- | --- | --- | --- | --- | --- | --- | --- | --- | --- | --- |
| Set5 | 28.44 | 0.1517 | 27.08 | 28.96 | 0.0484 | 0.0348 | 27.95 | 29.98 | 0.0527 | 0.0472 |
| Set14 | 26.10 | 0.1932 | 24.72 | 26.47 | 0.0764 | 0.0697 | 25.80 | 27.17 | 0.0765 | 0.0793 |
| BSD100 | 25.90 | 0.2323 | 24.19 | 25.92 | 0.0936 | 0.0935 | 24.93 | 26.27 | 0.0893 | 0.1012 |

TABLE III: Average PSNR and PS before and after fine-tuning ENet (Section V-A and VSRGAN (Section V-B) for the problem of artifact removal.

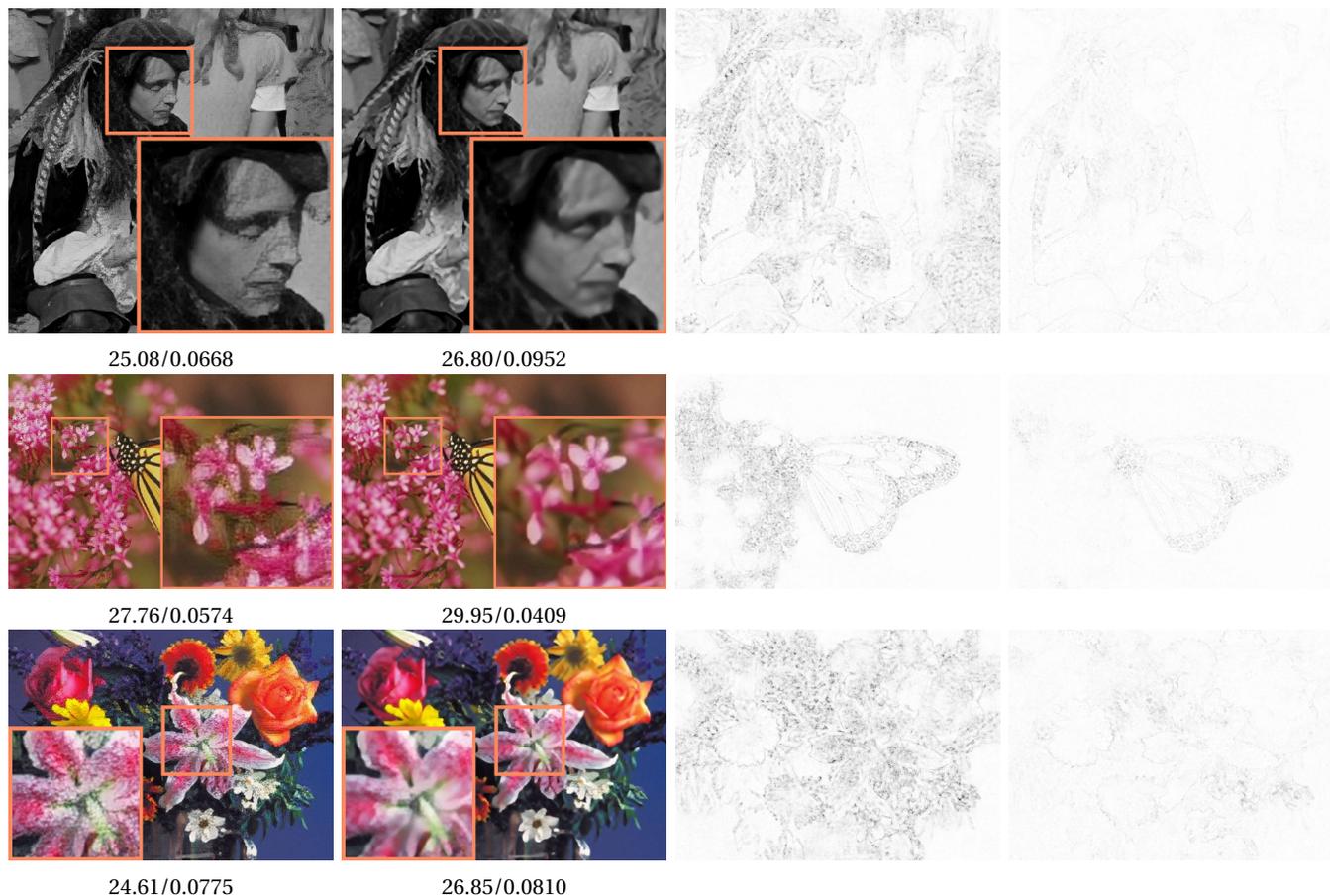

**Figure 9:** Correcting artifacts produced by VSRGAN (Section V-B.) The first column shows the VSRGAN [11] output, the second columns shows the fine-tuned VSRGAN∗ output. The third and fourth columns show the corresponding variance maps for VSRGAN and VSRGAN∗, respectively (for visualization purposes, we show high variance points in black pixels, whereas low variance points are encoded in white).

We must emphasize here that the applicability of our proposed method is not limited to the experiments described in this paper. Indeed, our fine-tuning algorithm for image enhancement is agnostic to the inverse imaging problem at hand, the CNN model, or the types of artifacts encountered or their cause. For example, the artifacts generated by ENet and those generated by VSRGAN seen in Section V were vastly different in their cause and nature. Indeed, ENet's artifacts were hypothesized to be the result of learning a recurring texture between HR and LR patches in the training dataset ([13]), whereas we showed in Section V-B1 that VSRGAN's artifacts are due to high uncertainty levels associated with a complex generative training. Similarly, inspecting SR images of CNNs trained for another scale factor in Section IV-A revealed a diverse range of artifacts, from a minor change in color space (e.g. EDSR), to significant distortions at the output (e.g. VSRGAN). Despite the diversity of artifacts, our fine-tuning algorithm was successful at enhancing the output in all cases, as long as the observed artifacts are not severe (see Section VI-E). Furthermore, while we presented our fine-tuning approach in its most elementary form, more evolved variants of our proposed method may be easily derived to better suit other imaging problems. The medical imaging community, for example, may want to incorporate additional regularizers in Equation 1 to enforce prior knowledge of the image's expected characteristics, for example enforcing a sparse representation in a given domain. Additionally, if prior knowledge regarding the location of the GAN artifacts is available, for example using the variance maps derived in Section V-B1, such knowledge can be used to employ a masked formulation of our proposed fine-tuning method and obtain improved results. Finally, we remark that while our experiments on artifacts due to Generative Adversarial Networks focused on the ENet and VSRGAN models, our method is model agnostic and thus can be generalized to any GAN-based model which produces mild artifacts in the SR solution.

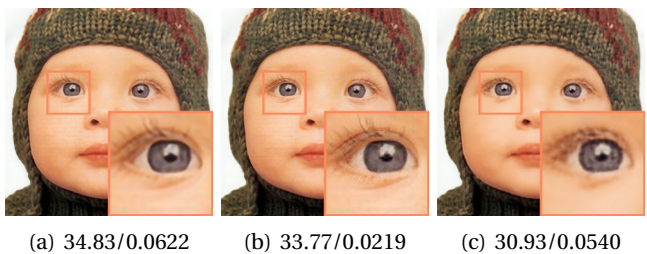

(a) 34.83/0.0622  (b) 33.77/0.0219  (c) 30.93/0.0540

**Figure 10:** Comparing results obtained from our fine-tuned models (a) VSRMSE∗ and (b) ENet∗ with (c) DIP [9] (Section IV-A). All these models were optimized to learn SR for scale factor 3.

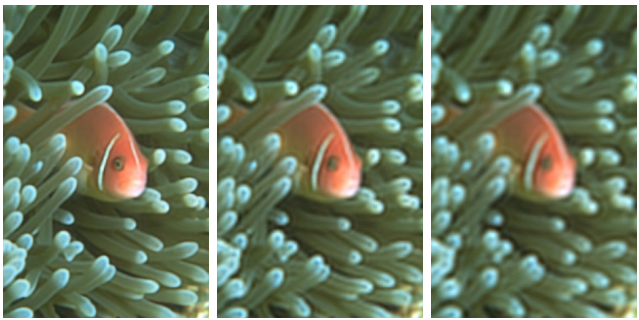

**Figure 11:** Fine-tuning EDSR for super-resolving LR images that have undergone circular blur of radii $r = 3, 5$ and $7$, respectively. The larger the blur, the more the perceptual quality of EDSR∗ decreases.

### B. Comparison with internal learning methods

As our framework utilizes a form of internal learning, it is natural to ask what makes our framework different from ZSSR [6] or DIP [9], two popular internal learning-based methods in today's literature. The most significant difference lies in our respective objectives, as ZSSR and DIP's end-goal is to propose a new reconstruction from scratch for each LR image encountered at test time, while our internal learning was introduced with the purpose of domain adaptation for artifact correction *of an existing model*. This novel set-up allows us to take advantage of the prior learning resulting from earlier training on large datasets. In Table IV we compare the results obtained when running the DIP code available at https://github.com/DmitryUlyanov/deep-image-prior for learning super-resolution for scale factors 2 and 3. We compare their results with the SR solutions learned by our fine-tuned VSRMSE∗, ENet∗, and EDSR∗ models. From the reported metrics we observe that in most cases our fine-tuned models quantitatively surpass the DIP method. This is consistent with our earlier observation that using pre-trained CNNs for internal learning as opposed to random networks provides us with a stronger prior. In Figure 10, we compare the output of two of our fine-tuned models, VSRMSE∗ and ENet∗ with the SR obtained from DIP. From this figure we observe that both of our fine-tuned models produce results of high perceptual quality. Furthermore, DIP's SR output appears to contain a small amount of noise that is not observed in our models. An additional property that separates our proposed framework from ZSSR or DIP is how easily our method can be integrated in an already existing software. This is particularly relevant in industry whose products may rely on a CNN trained for a task that may be modified for various reasons, for example when faced with a new SR task or LR input domain. Implementing our method in such a product requires very relatively few steps, as shown in the pseudocode in Algorithm 1, and thus can be implemented as a simple "wrapper" code around any existing SR method. In Section IV-B, we noted that fine-tuning EDSR to obtain EDSR∗ resulted in a similar quantitative performance as ZSSR. However, the implementation of fine-tuning for EDSR∗ was more straight-forward than adapting the code made available by ZSSR to our computational environment, as a working environment for EDSR was already available to us. Thus in applications in which time is a constraint, fine-tuning a model with our proposed framework may be preferable.

Finally, we emphasize that internal learning methods such as ZSSR and DIP would not be capable of reducing GAN artifacts as shown Section V. This is because their framework is specific to the task of solving an inverse imaging problem from scratch, while our fine-tuning method serves as a general-purpose method capable of enhancing SR CNNs for a variety of tasks.

### C. Comparison with SR CNNs trained with the correct degradation operator

It is expected that a network fine-tuned on internal data to adapt to $A_{test}$ will not perform as well as the same CNN $f_\theta(\cdot)$ trained on a large amount of low-resolution and high-resolution pairs generated with the correct $A_{test}$. For example if we had access to an EDSR network trained on a non-ideal LR dataset as described in Section IV-B, the resulting results would naturally surpass the EDSR∗ or ZSSR results presented in Table II. However, we note that such an EDSR network (or any other network) would still not generalize to any new SR setting $A_{test}$ and fail at generating pleasing SR solutions in such unseen cases. With the internal learning property of our framework, our fine-tuning method can adapt a fixed pre-trained network to new SR settings at test time. Of course this comes with the limitation that the algorithm needs to be executed again for each new test image and observation model.

| Dataset | Factor | DIP [9] | | VSRMSE∗ | | ENet∗ | | EDSR∗ | |
|---------|--------|---------|------|---------|------|-------|------|-------|------|
| | | PSNR | PS | PSNR | PS | PSNR | PS | PSNR | PS |
| Set5 | 2 | 33.29 | 0.0242 | 34.68 | 0.0337 | 35.49 | 0.0116 | 36.92 | 0.0137 |
| | 3 | 28.92 | 0.0369 | 32.56 | 0.0507 | 31.48 | 0.0299 | N/A | N/A |
| Set14 | 2 | 30.18 | 0.0388 | 31.09 | 0.0556 | 31.82 | 0.0279 | 33.00 | 0.0288 |
| | 3 | 26.37 | 0.0598 | 29.12 | 0.0902 | 28.38 | 0.0500 | N/A | N/A |
| BSD100 | 2 | 29.41 | 0.0442 | 30.01 | 0.0842 | 32.13 | 0.0461 | 31.69 | 0.0486 |
| | 3 | 26.17 | 0.0679 | 27.81 | 0.0868 | 28.07 | 0.1313 | N/A | N/A |

TABLE IV: Comparing our fine-tuned models with the solution obtained by DIP [9] for the disagreement in scale experiment (Section IV-A). Note that while producing superior PSNR performance than DIP, VSRMSE shows higher PS values than DIP. As VSRMSE was trained on videos as opposed to images, applying it in a still-image setting results in a slight decrease in performance.

| Dataset | Factor | Bicubic | | VSRMSE†[11] | | VSRMSE∗ | | EDSR†[3] | | EDSR∗ | |
|---------|--------|---------|------|-------------|------|---------|------|----------|------|-------|------|
| | | PSNR | PS | PSNR | PS | PSNR | PS | PSNR | PS | PSNR | PS |
| Set5 | 2 | 33.69 | 0.0479 | 36.24 | 0.0103 | 34.68 | 0.0337 | 38.04 | 0.0129 | 36.92 | 0.0137 |
| | 3 | 30.44 | 0.1021 | 32.21 | 0.0334 | 32.56 | 0.0507 | N/A | N/A | N/A | N/A |
| Set14 | 2 | 30.35 | 0.0676 | 32.12 | 0.0190 | 31.09 | 0.0556 | 33.72 | 0.0261 | 33.00 | 0.0288 |
| | 3 | 27.63 | 0.1370 | 28.97 | 0.0612 | 29.12 | 0.0902 | N/A | N/A | N/A | N/A |
| BSD100 | 2 | 29.49 | 0.0961 | 30.55 | 0.0300 | 30.01 | 0.0842 | 32.13 | 0.0461 | 31.69 | 0.0486 |
| | 3 | 27.14 | 0.1754 | 27.81 | 0.0868 | 28.07 | 0.1313 | N/A | N/A | N/A | N/A |

TABLE V: Comparing our fine-tuned model VSRMSE∗ and EDSR∗ with the corresponding CNNs trained on the correct $A_{test}$, denoted as VSRMSE†[11] and EDSR†[3].

For the sake of comparison, we obtain the VSRMSE [11] and EDSR [3] models trained for scale factors 2 and 3 and run these models to super-resolve LR images at these scale factors. We refer to these models as VSRMSE† and EDSR†, respectively. We quantitatively compare these with the fine-tuned VSRMSE∗ and EDSR∗ that were never trained for such scale factors but instead fine-tuned as explained in Section IV-A. The results in Table V are consistent with our earlier argument that networks trained on the correct $A_{test}$ are guaranteed to perform better than fine-tuned networks. Again, we re-iterate that our method's objective is not to produce a new state-of-the-art network, but instead is meant to be used as a tool that enhances or corrects results of already available models.

### D. Running times

In Table VI we show the running times of our fine-tuning method for the experiments described in Section IV and V. We observe that fine-tuning for EDSR∗ results in the fastest runtimes. Because the EDSR architecture performs all of its convolution operations in low-resolution space, faster forward passes ensue which naturally lead to a shorter optimization process. Conversely, fine-tuning the VSRMSE network is slower as the convolutions are applied in the high-resolution space.

Furthermore, it is clear that the runtime of our algorithm depends on how far from the ideal solution $\psi$ our original $\psi_0$ lies in parameter-space. We have seen in Section IV-A that fine-tuning the EDSR network for scale factor disagreements simply amounted to mapping to the correct color space. A smaller number of iterations are required for such a simple task. The VSRGAN artifacts, however, are much stronger and a considerable amount of learning is necessary to correct for them, thus the resulting longer computation times. Similarly note that when removing artifacts due to GANs as described in Section V, the fine-tuning algorithm converges in no more than a couple of minutes regardless of the model employed. That is because only a small amount of pixels in the initial solution $f_{\hat{\theta}}$ need to be changed.

### E. Limitations

While our fine-tuning framework presents many assets detailed above, some limitations remain. One major limitation is that the more severe the artifacts present in the original $f_\theta$, the less likely it is that our optimization will successfully converge to a pleasing solution. Consider for example the original solution provided by EDSR for super-resolving at scale factor 3 when trained for factor 4 (shown in the second row of Figure 4). As explained in Section IV-A, the observed strong artifacts are the result of inserting a randomly initialized convolution layer adapted to the scale factor 3 problem in the EDSR architecture. In this case, our experiments have found that our fine-tuning algorithm can *still* converge to a perceptually pleasing solution, however, it does so in a significantly longer time and, in addition, requires special care in hyper-parameter tuning and regularization. This defeats the original motivation for introducing our method. Similarly if presented with a GAN network that produces major artifacts in its super-resolved output, it is unlikely that the straightforward application of Equation 2 will succeed.

|  | Section IV-A | | | | | | | | | Section IV-B | | | Section V | | | | |
|---|---|---|---|---|---|---|---|---|---|---|---|---|---|---|---|---|---|
| | EDSR | | | ENet | | | VSRMSE | | | EDSR | | | ENet | | | VSRGAN | |
| Set5 | Set14 | BSD | Set5 | Set14 | BSD | Set5 | Set14 | BSD | BSD(a) | BSD(b) | Set5 | Set14 | BSD | Set5 | Set14 | BSD |
| 0.9 | 1.2 | 0.5 | 4.6 | 5.7 | 3.3 | 9.7 | 13.7 | 8.3 | 0.7 | 1.3 | 1.2 | 1.3 | 1.1 | 2.0 | 3.3 | 2.2 |
| 910 | 570 | 370 | 3290 | 2282 | 1905 | 2880 | 2225 | 2029 | 886 | 1544 | 880 | 442 | 340 | 180 | 193 | 151 |

TABLE VI: Average processing times in minutes (top row) and corresponding number of iterations (bottom row). These experiments were performed on a GTX 1080 GPU card. BSD(a) refers to the blind setting and BSD(b) refers to the nonblind setting.

Furthermore, it should be clear that our proposed fine-tuning framework in Equation 1 requires the new degradation operators $A_{test}$ to be similar in nature to the A assumed in training. Indeed our method is not expected to be robust to degradation operators that differ by a large amount, or if the amount of the new degradation is too strong. In Figure 11 we show the result of applying successively higher amount of blurs on an input image and super-resolving it by applying Equation 1 on EDSR. Clearly as higher levels of blur are applied, the less perceptually satisfying the solution provided by EDSR∗ becomes. Moreover notice that in Section IV-A, we have constrained our experiments to address the case of fine-tuning SR CNNs that were trained for scale factor 4 for new problem of scale factors 2 and 3. We did not consider correcting for other unseen factors such as 5, 6, 7 or 8. Indeed our experiments found that our fine-tuning method was most successful when the factor for which the network was trained for was higher than the factor for which we fine-tune it. This is expected, as in such cases the solution $f_\theta(\cdot)$ was trained for a much less complex problem than the new one it is presented to at test time and hence cannot provide leverage when used as an initial solution. In these cases, better results will be obtained by reverting to a method that fully relies on internal learning from randomized parameters, such as ZSSR [6] or DIP [9] or by training a deep CNN with the appropriate $A_{test}$.

Finally, our proposed framework still presents room for improvement in that it is not fully automated and may still requires some amount of hyper-tuning for a given experiment, e.g., to select the ideal learning rate and/or gradient descent algorithm, or select the optimal stopping criterion. However, it is clear that the amount of work associated with these hyper-parameter decisions is minuscule compared to those required for re-training a neural network to obtain the appropriate solution.

## VII. CONCLUSION

In this work, we proposed a fine-tuning algorithm to correct for the solution learned by SR CNNs under various scenarios. We achieve this by fine-tuning the parameters of several neural networks to satisfy a data-fidelity term given a new test data point. Our method is fully self-supervised and distinguishes itself in that it does not necessitate large training datasets or evolved training experiments to be effective. We applied our algorithm to the case of adapting to a new SR setting at test time (Section IV) and correcting hallucinated GAN content (Section V). In each of our experiments, our approach consistently resulted in images of higher qualitative and quantitative quality.

Such a generic framework for post-processing the solutions of CNNs using fine-tuning on a test input image has to the best of our knowledge not been explored in the literature, other than in our initial work presented in [1]. The novelty of our approach naturally provides room for further research and improvement on our proposed framework. An example of such improvements may be obtained by combining the abundant research in analytical methods for inverse imaging problems with efficient fine-tuning in pre-trained parameter space. One could also easily introduce additional regularizers to the proposed framework, which may be particularly beneficial to communities such as the medical imaging community. Investigating further in such directions will provide deep learning scientists with more leeway to address some of the shortcomings of deep learning. Indeed, through the application of reliable, well-established analytical approaches offered by the image processing community to this framework, one may be one step closer to achieving restorations of previously unseen quality produced by DNNs for inverse imaging problems.